\documentclass[a4paper]{jpconf}
\usepackage{graphicx}

\usepackage[square,sort,comma,numbers]{natbib}

\begin{document}
\title{Status of the Project 8 Phase II}

\author{Mathieu~Guigue, on behalf of the Project 8 collaboration}
\address{Pacific Northwest National Laboratory, Richland, WA, USA}
\ead{mathieu.guigue@pnnl.gov}

\begin{abstract}
The Project 8 collaboration aims to measure the absolute neutrino mass scale using cyclotron radiation emission spectroscopy on the beta decay of tritium.
The second phase of the project will measure a continuous spectrum of molecular tritium beta decays and extract the tritium endpoint value with an eV or sub-eV scale precision.
Monoenergetic electrons emitted by gaseous $^{83\mathrm{m}}$Kr atoms are used to determine the relationship between cyclotron frequency and electron energy.
This study allows us to optimize both the event reconstruction algorithm and the hardware configuration, in preparation for measuring the tritium beta decay spectrum.
Phase II will benefit from a gas system of krypton and tritium that will allow measurement of and offline correction for magnetic field fluctuations.
We present the recent progress in understanding the electron kinematics and implementing the magnetic field calibration.
\end{abstract}

\section{Toward a neutrino mass measurement}

The measurement of neutrino flavor oscillations has demonstrated that neutrinos have a non-zero mass.
This discovery leads to questioning the origin and absolute scale of their masses.
A way of directly measuring this scale is to study the tritium beta spectrum.
Near the endpoint, the spectrum shape $\frac{\mathrm{d}N}{\mathrm{d}E_{\mathrm{kin}}}$ depends on the neutrino mass $m_{\beta}$ as 
\begin{equation}
\frac{\mathrm{d}N}{\mathrm{d}E_{\mathrm{kin}}}\propto (Q-E_{\mathrm{kin}})\sqrt{(Q-E_{\mathrm{kin}})^2-m_{\beta}^2/c^2},
\end{equation}
where $Q$ is the endpoint value (of about $18.6~\mathrm{keV}$) and $E_{\mathrm{kin}}$ the electron kinetic energy.

The Project 8 collaboration is taking a four-phase approach \cite{Ashtari2017} to measure the tritium beta spectrum and reach a sensitivity to the neutrino mass of $40~\mathrm{meV}$.
The Cyclotron Radiation Emission Spectroscopy (CRES) technique, the soundness of which has been demonstrated by the Collaboration \cite{Asner2015}, measures the cyclotron frequency of magnetically trapped electrons that relates to the electron kinetic energy as
\begin{equation}\label{eq:cyclotron-frequency}
f_c = \frac{1}{2\pi}\frac{eB}{m_e+E_{\mathrm{kin}}/c^2},
\end{equation}
where $e$ and $m_e$ are the charge and mass of the electron and $B$ is the magnetic field experienced by the electron.
Conversion electrons at energies of $17.8~\mathrm{keV}$, $30.4~\mathrm{keV}$ and $32~\mathrm{keV}$ and a gaseous nature similar to tritium makes $^{83m}$Kr an appropriate source to measure the magnetic field $B$ and calibrate the instrument.

\section{Phase II instrumental setup}

\begin{figure}[t]
\centering
\includegraphics[width=0.8\textwidth]{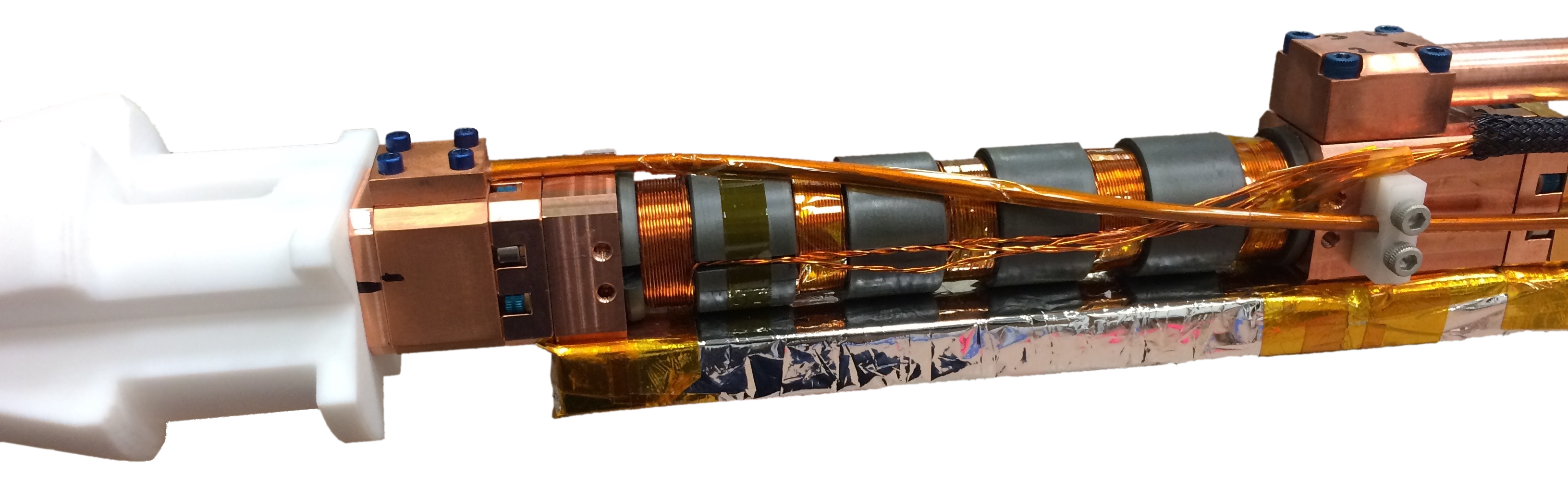}
\begin{minipage}[b]{0.59\textwidth}
\includegraphics[width=\textwidth]{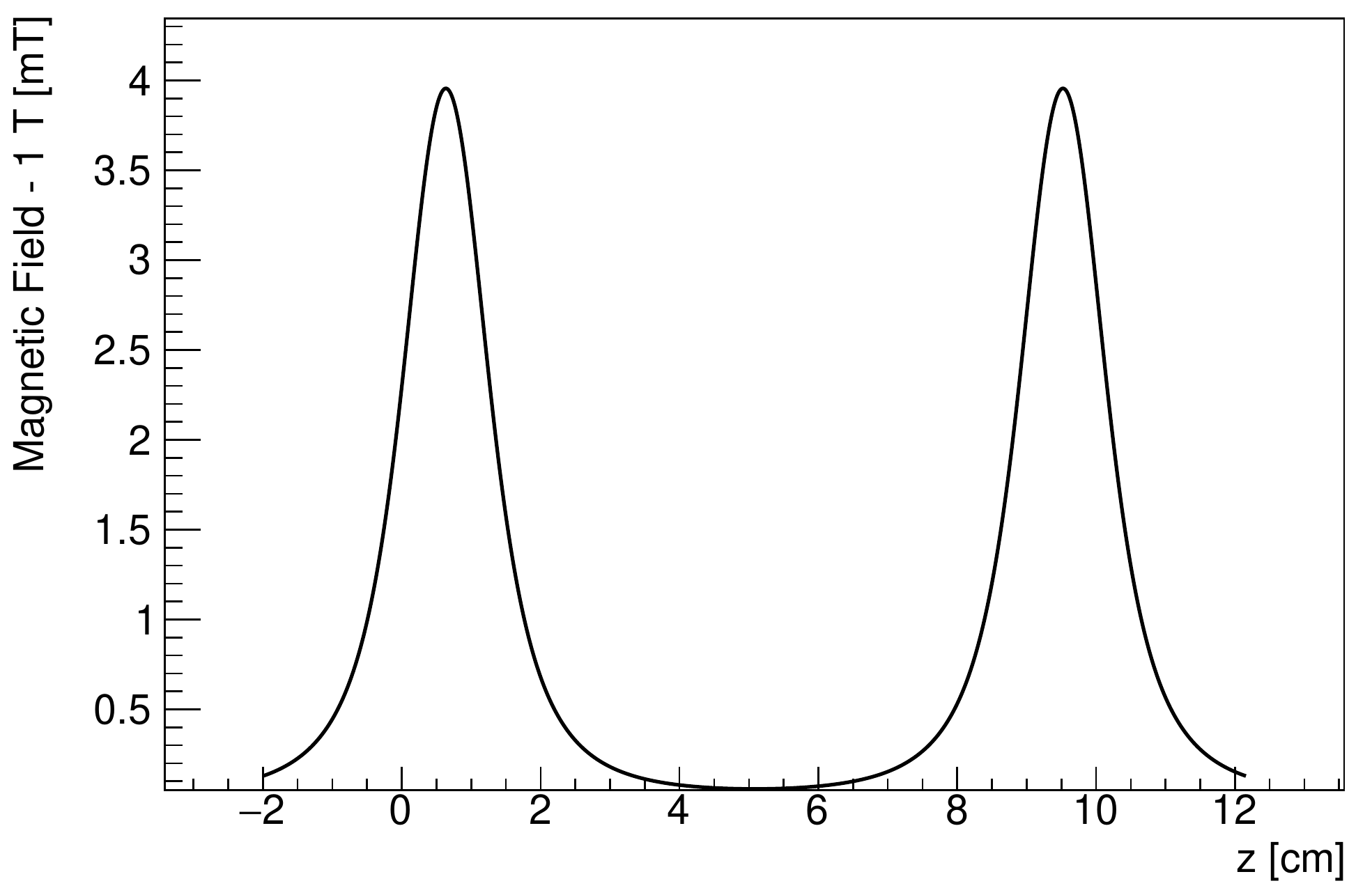}
\end{minipage}
\begin{minipage}[b]{0.4\textwidth}
\caption{\label{fig:cell-bathtub}Phase II instrumental setup. 
Top: a picture of the circular waveguide and the 5 trapping coils. 
The copper-plate reflector is placed between the plastic cone on the far left and the waveguide.
Bottom: The magnetic trap configuration called the ``bathtub''. 
A positive current is set in the most extreme coils to create a bump in the holding magnetic field.
Electrons generated within the two bumps can be trapped if their pitch angle is large enough.
}
\end{minipage}
\end{figure}

Phase II aims to make the first measurement of the continuous tritium beta spectrum using the CRES technique \cite{Pettus2017}.
A tritium compatible cell has been built for this purpose.
Figure \ref{fig:cell-bathtub} shows the circular waveguide containing the gas and the five trapping coils.
This waveguide is immersed in a $1~\mathrm{T}$ holding magnetic field along the waveguide axis: in this configuration the cyclotron frequency is about $26~\mathrm{GHz}$.
The trapping coils can produce several trapping configurations: among these, the bathtub one (depicted on Fig. \ref{fig:cell-bathtub}) corresponds to positive currents applied to the most external coils and is of interest as the corresponding electron trapping region has a larger magnetic uniformity.
Electrons that are emitted by the decaying atoms and magnetically trapped experience a cyclotron motion whose generated power is guided by the waveguide to the acquisition chain.
A reflector is used to reflect the power going down the waveguide back to the acquisition chain, increasing the total received power.

The cyclotron signal is amplified by a set of two low-noise amplifiers, downmixed and digitized.
Phase II makes use of two acquisition systems.
The first one is a Real-time Spectrum Analyzer (RSA) which is used for diagnostic and possesses a trigger based on the detected instantaneous power.
Its frequency range  is $85~\mathrm{MHz}$ and the central frequency can be moved over several GHz.
The second one is a ROACH2 digitizer, which allows measurement of three $100~\mathrm{MHz}$-wide windows.
This instrument currently records data in streaming mode, but a trigger mode is under development. 

\section{Event reconstruction and frequency corrections}

\begin{figure}[t]
\includegraphics[width=0.6\textwidth]{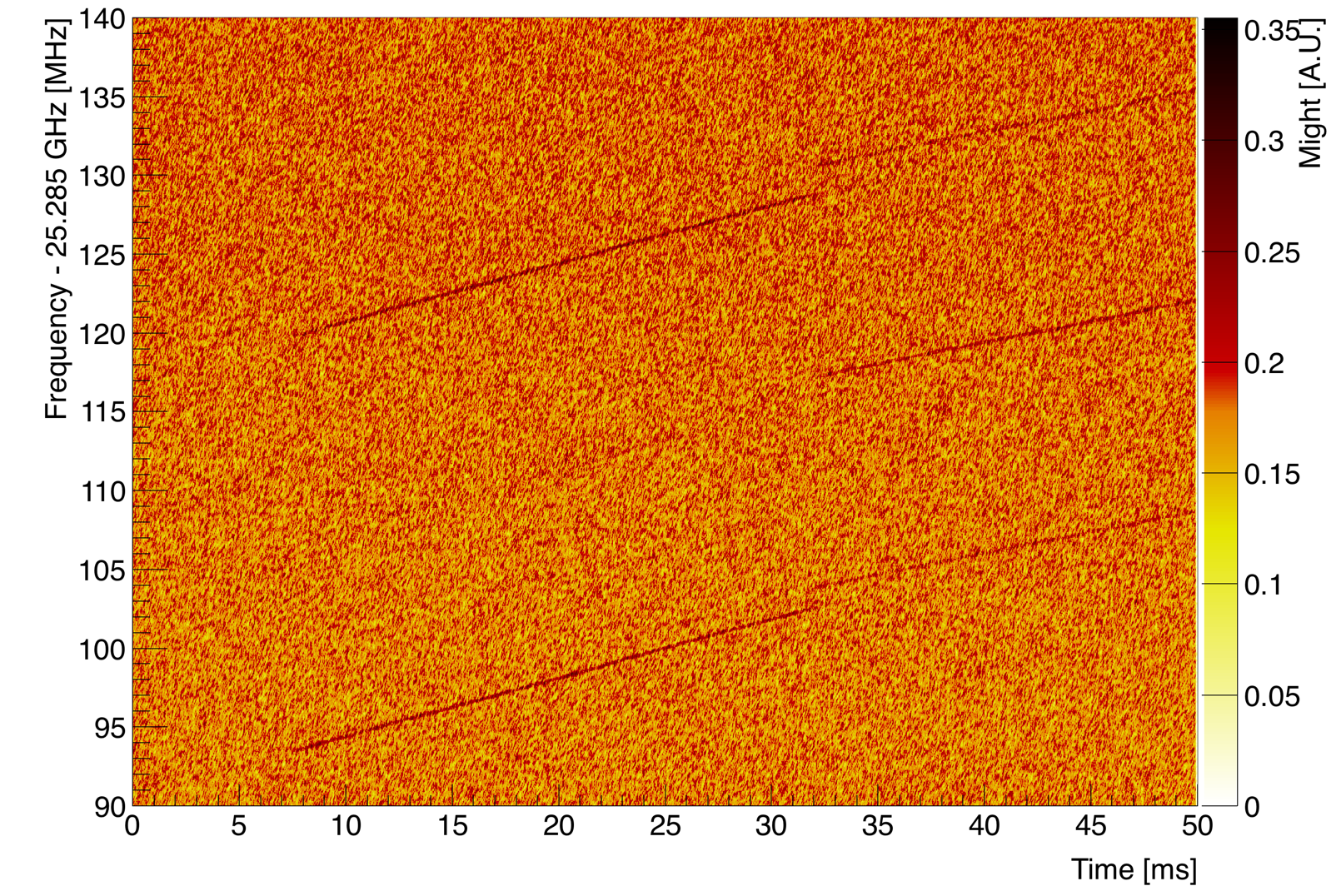}
\begin{minipage}[b]{0.4\textwidth}
\caption{\label{fig:spectrogram}Spectrogram of a $50~\mathrm{ms}$ signal record.
Power generated by a trapped electron shows up as lines on top of the RF noise background.}
\end{minipage}
\end{figure}
\begin{figure}[t]
\includegraphics[width=0.56\textwidth]{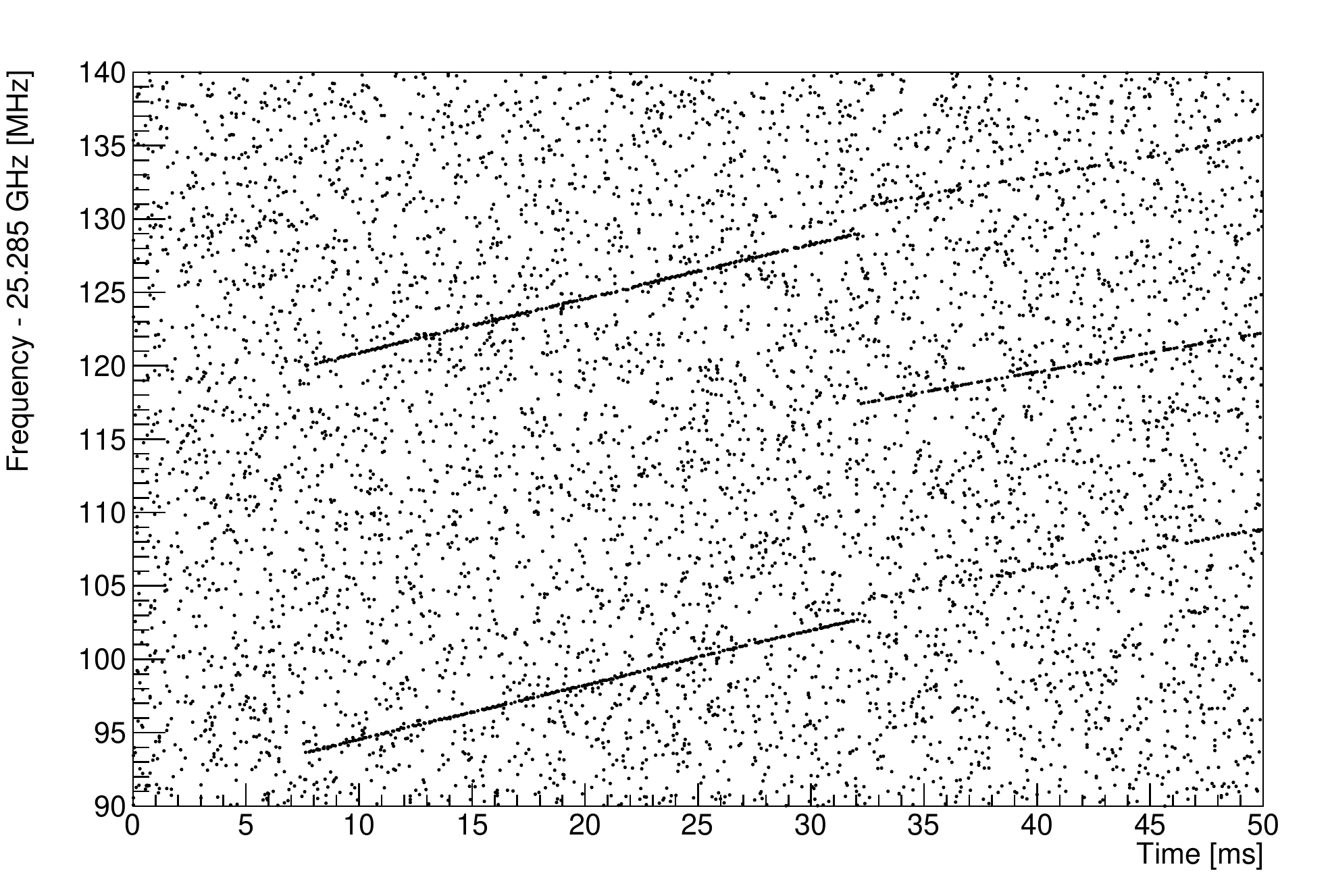}
\begin{minipage}[b]{0.4\textwidth}\caption{\label{fig:sparsespectrogram}Sparse spectrogram generated from Fig. \ref{fig:spectrogram} after applying a discriminator based on a threshold.
The high power and aligned dots correspond to the tracks.
Next step consists in clustering aligned points together to form the tracks and remove the noise.}
\end{minipage}
\end{figure}

From the recorded time series, spectrograms such as presented in Fig. \ref{fig:spectrogram} are produced.
Excess of power generated by the electrons forms tracks: as radiation is emitted and the electron loses power the radiation frequency increases with time, as shown by Eq. (\ref{eq:cyclotron-frequency}). 
A reconstruction algorithm is used to extract the starting energy of the electron.
The first step of this algorithm is to apply a offline trigger to only select the bins with high power: the result of this step for the spectrogram in Fig. \ref{fig:spectrogram} is depicted on Fig. \ref{fig:sparsespectrogram}.
The second step uses a clustering algorithm to find the alignment of power excess and remove noise.
For each cluster representing a track, the start frequency (among other quantities) is extracted.
The final step first combines parallel tracks together.
Then a set of parallel tracks that ends at the same time as another one starts are finally grouped together into an event and the event start frequency is recorded.
The extracted frequency corresponds to the initial frequency of an electron, and the starting energy can be obtained provided the magnetic field is known.

\begin{figure}[t]
\includegraphics[width=0.6\textwidth]{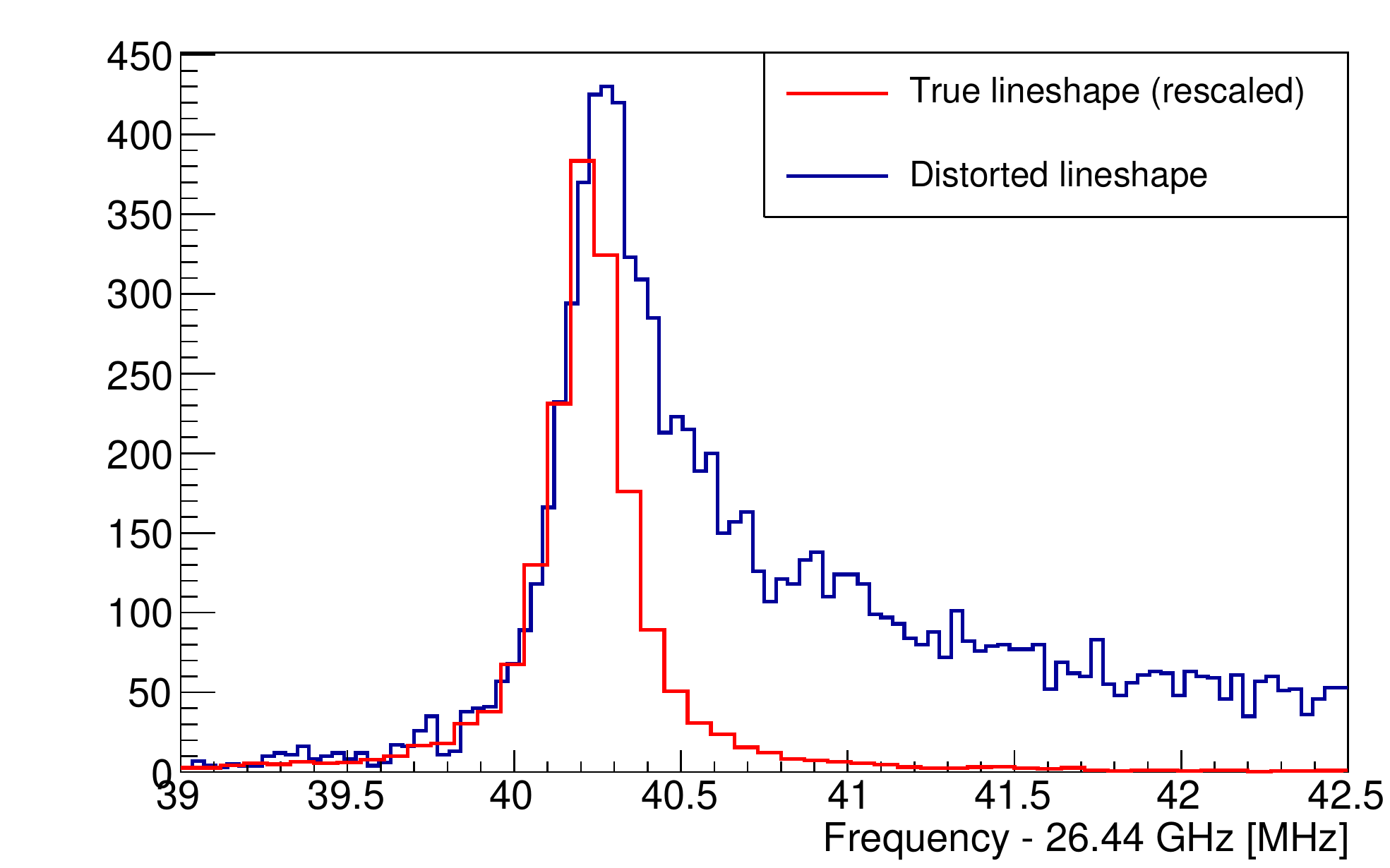}
\begin{minipage}[b]{0.4\textwidth}
\caption{\label{fig:lineshape-distortion}Simulation of a $30~\mathrm{keV}$ line (red) and its distortion (blue) induced by a characteristic bathtub trap.
The distorted line has a higher average frequency. 
The high frequency tail corresponds to electrons with low pitch angles that explore higher magnetic field regions of the trap.
}
\end{minipage}
\end{figure}

The difficulty in converting the measured start frequency into an energy is that the magnetic field experienced by the electron depends on the electron's directionality: we call the angle between the holding magnetic field direction and the electron velocity the pitch angle $\theta$.
The measured start frequency goes as
\begin{equation}\label{eq:distorted-frequency}
f_s= \frac{1}{2\pi}\frac{e(B+b(\theta ))}{m_e+E_{\mathrm{kin}}/c^2},
\end{equation}
where $b(\theta )$ is the contribution induced by the presence of the magnetic trap and its exploration by the electron.
As low pitch angle electrons will experience a higher averaged magnetic field value, this will cause a distortion of the spectrum toward high frequencies.
Such distortion of a $30~\mathrm{keV}$ Krypton line induced by a bathtub trap configuration is shown on Fig. \ref{fig:lineshape-distortion}.
Moreover this causes a high frequency tail.
If one can determine the pitch angle value for each detected electron, one could correct for this distortion using Eq. (\ref{eq:distorted-frequency}). 

\begin{figure}[t]
\includegraphics[width=\textwidth]{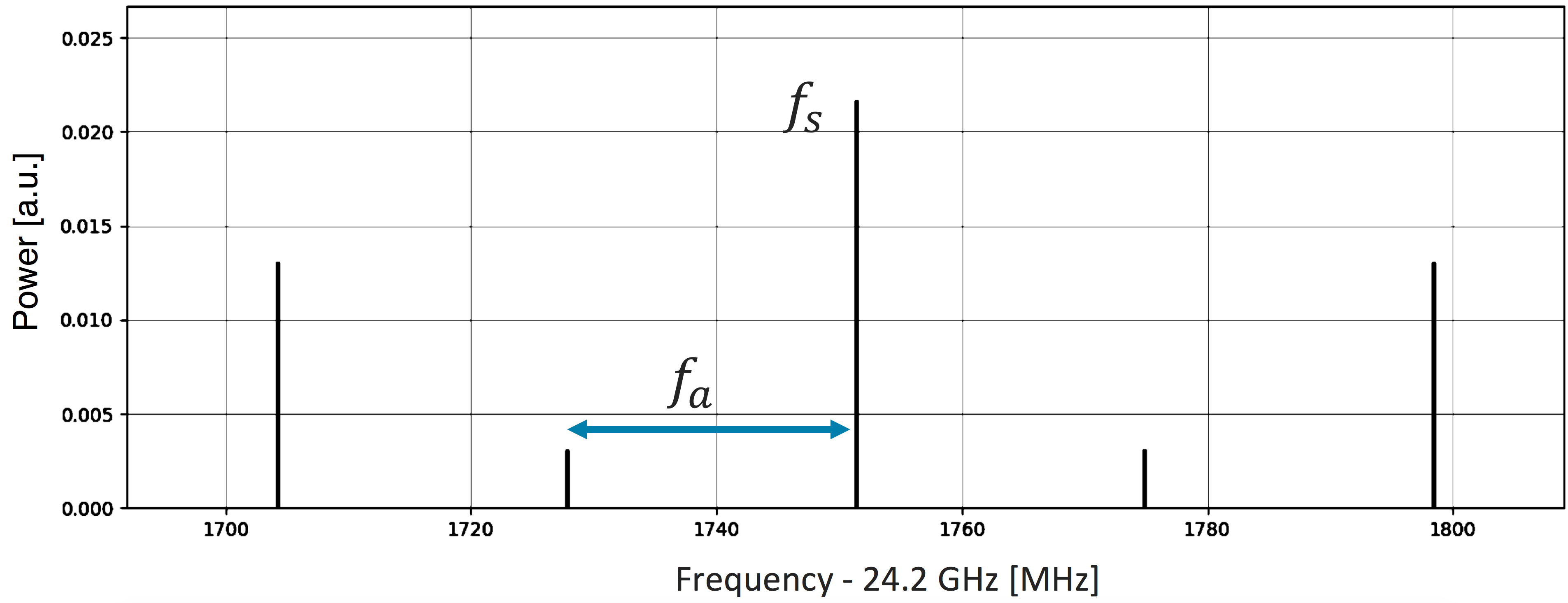}
\caption{\label{fig:comb-spectrum}Illustration of the comb spectrum.
The start frequency $f_s$ of the event is defined by the position of the main or ``central peak''. 
Sidebands symmetrically surround the main peak and the frequency distance between each peak is equal to the axial frequency.
The first order sidebands are $f_a$ away from the main peak, the second order sidebands are $2f_a$ away from the main peak and so on. }
\end{figure}

It turns out that several characteristics of our detected events contain some information about the pitch angle value.
One of them is the frequency of the electron motion along the waveguide axis called {\it axial frequency}.
In the case of a bathtub trap configuration, the axial frequency can be calculated using
\begin{equation}
f_a = \frac{1}{2\pi}\frac{\sqrt{\frac{2E_{\mathrm{kin}}c^2}{m_e}}}{\frac{2\pi L_0}{\sin\theta}+\frac{2L_1}{\cos\theta}},
\end{equation}
where $L_1$ defines the size of the flat part of the trap and $L_0$ is a measurement of the curvature of the edges of the trap.
This phenomenon is responsible for a comb structure of the instantaneous power spectrum, as represented on Fig. \ref{fig:comb-spectrum}.
The axial frequency corresponds to the distance between parallel tracks; more precisely, it is the frequency distance between the main carrier and the sideband. 
A consequence of this phenomenon on the spectrogram as seen on Figs. \ref{fig:spectrogram} and \ref{fig:sparsespectrogram} is that an event could possesses several parallel tracks spaced by the the axial frequency.
From the measurement of this distance and a study of its distribution, one can extract the pitch angle for each event and correct the measured start frequency using (\ref{eq:distorted-frequency}) and a model describing the magnetic trap.
An analysis using the sidebands to correct for the effect of the trap on the Krypton lineshape for Phase II is currently under development.

\begin{figure}[t]
\includegraphics[width=0.6\textwidth]{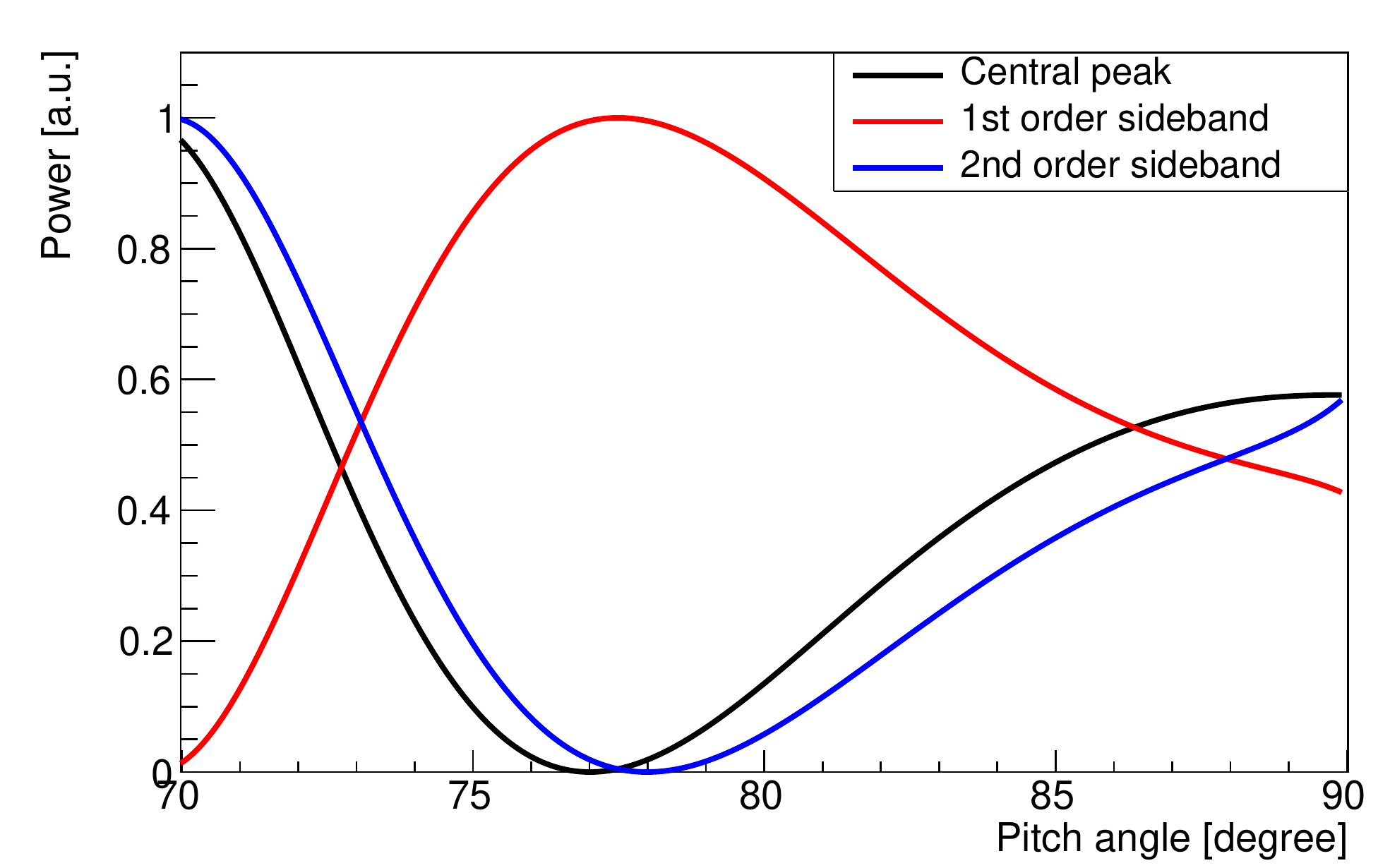}
\begin{minipage}[b]{0.4\textwidth}
\caption{\label{fig:power-peaks}Power of the main peak (black) and first (red) and second order (blue) sidebands as a function of the electron pitch angle.
The power of each peak is given in arbitrary units.
We can see that the pitch angle controls the phase of the interference pattern that cancels the peaks.
}
\end{minipage}
\end{figure}

Figures \ref{fig:spectrogram} and \ref{fig:sparsespectrogram} also show a difference in the main peak power to sideband ratio between the first and second sets of parallel tracks.
This effect can be explained by the presence of the waveguide reflector used to reflect half of the cyclotron power back to the acquisition chain.
Because of this reflector, the detected signal is a superposition of two components.
Depending on the distance of the electron to the reflector or its pitch angle, this interference can be either constructive or destructive.
Figure \ref{fig:power-peaks} shows the power for the main peak and the first and second order sidebands as a function of the pitch angle.
When the power of the main peak and even-order sidebands is low, the power of odd-order peak is maximal.
This phenomenon explains why some tracks may completely vanish and reappear upon scattering when the electron pitch angle changes.


The event depicted on Fig. \ref{fig:sparsespectrogram} presents many interesting features which are characteristic to the Phase II implementation of the CRES.
We have two sets of parallel tracks that are generated by a single electron.
The first set of parallel tracks are sidebands: the electron pitch angle is such that the main peak is cancelled by the wave reflected by the reflector.
After some time, the electron collides with an atom from the background gas and its pitch angle changes such that the main peak reappears and the sidebands become less powerful.
As the electron has collided, it also loses some energy increasing the start frequency of the second set of tracks.

\section{Conclusion and perspectives}

Phase II of the Project 8 experiment is currently taking data and the analysis is in progress.
The collaboration has developed a phenomenological model thats explains the structure of our events and some features induced by the instrument configuration.
A robust algorithm is being implemented for correcting the lineshape distortions and improving the instrument's energy resolution using sidebands.
Magnetic field calibration is under way in preparation to connect a tritium source and record our first differential tritium spectrum.

\bibliography{iopart-num}

\providecommand{\newblock}{}
\begin{thebibliography}{1}
\expandafter\ifx\csname url\endcsname\relax
  \def\url#1{{\tt #1}}\fi
\expandafter\ifx\csname urlprefix\endcsname\relax\def\urlprefix{URL }\fi
\providecommand{\eprint}[2][]{\url{#2}}

\bibitem{Ashtari2017}
{Ashtari Esfahani} A~{\etal} (Project 8) 2017 {\em J. Phys. G: Nucl. Part.
  Phys.\/} {\bf 9} 054004

\bibitem{Asner2015}
Asner D~M~{\etal} (Project 8) 2015 {\em Phys. Rev. Lett.\/} {\bf 114} 162501

\bibitem{Pettus2017}
Pettus W (Project 8) 2017 {\em Proc. XV TAUP\/}

\end{thebibliography}

\end{document}